\documentclass{PoS}
\usepackage{amsmath}

\title{Emergent quantum geometry from stochastic random matrices%
\footnote{Report No.: KUNS-2809}}

\ShortTitle{Emergent quantum geometry from stochastic random matrices}

\author{\speaker{Masafumi Fukuma} and Nobuyuki Matsumoto\\
Department of Physics, Kyoto University\\
Kyoto 606-8502, Japan\\
E-mail: \email{fukuma@gauge.scphys.kyoto-u.ac.jp}, 
\email{nobu.m@gauge.scphys.kyoto-u.ac.jp}
}

\abstract{
Towards formulating quantum gravity, 
we present a novel mechanism for the emergence of spacetime geometry from randomness. 
In [arXiv:1705.06097], 
we defined for a given Markov stochastic process 
``the distance between configurations,'' 
which enumerates the difficulty of transition between configurations. 
In this article, we consider stochastic processes of large-$N$ matrix models, 
where we regard the eigenvalues as spacetime coordinates. 
We investigate the distance for the effective stochastic process of one-eigenvalue, 
and argue that this distance 
can be interpreted in noncritical string theory 
as probing a classical geometry with a D-instanton. 
We further give an evidence that, 
when we apply our formalism to a tempered stochastic process of $U(N)$ matrix, 
where the 't Hooft coupling is treated as another dynamical variable, 
a Euclidean AdS$_2$ geometry emerges in the extended configuration space 
in the large-$N$ limit, 
and the horizon corresponds to the Gross-Witten-Wadia phase transition point. 
}

\FullConference{Corfu Summer Institute 2019 "School and Workshops on Elementary Particle Physics and Gravity" (CORFU2019)\\
		31 August - 25 September 2019\\
		Corfu, Greece}

\begin{document}

%%%%%%%%%%%%%%%%%%%%%%%%%%%%%%%%%%%%%%%
%%%%%%%%%%%%%%%%%%%%%%%%%%%%%%%%%%%%%%%
\section{Introduction}
\label{sec:introduction}
%%%%%%%%%%%%%%%%%%%%%%%%%%%%%%%%%%%%%%%
%%%%%%%%%%%%%%%%%%%%%%%%%%%%%%%%%%%%%%%

There has long been an expectation that 
quantum mechanics has its origin in randomness 
\cite{Nelson:1966sp,Parisi:1980ys}. 
There also have been attempts to explain 
the observed values of physical constants 
by making the constants random variables 
\cite{Froggatt:1978nt,Froggatt:1995rt,Coleman:1988cy,Coleman:1988tj}.
A natural question we may then have is 
whether a quantum theory of gravity can be formulated 
based on randomness. 
In order for such framework to be meaningful, 
there must be a mechanism for 
the {\it emergence of spacetime geometry from randomness}.

In \cite{FMU}, 
we defined a geometry for an arbitrary Markov stochastic process 
such that the distance between configurations quantifies 
the difficulty of transition between them. 
In this article, 
we discuss that the above mechanism can be realized 
by considering a stochastic process of matrices, 
where we regard the eigenvalues as coordinates of a spacetime. 
We investigate the distance for the effective stochastic process of one-eigenvalue, 
and argue that this distance 
can be interpreted in noncritical string theory 
as probing a classical geometry with a D-instanton. 
We further give a preliminary result that, when we apply our formalism to 
a stochastic process of $U(N)$ matrix 
with treating the 't Hooft coupling as another dynamical variable, 
a Euclidean AdS$_2$ geometry emerges in the extended configuration space 
in the large-$N$ limit, 
where the horizon corresponds to the Gross-Witten-Wadia phase transition point. 

%%%%%%%%%%%%%%%%%%%%%%%%%%%%%%%%%%%%%%%
%%%%%%%%%%%%%%%%%%%%%%%%%%%%%%%%%%%%%%%
\section{Distance between configurations in a Markov stochastic process}
\label{sec:model}
%%%%%%%%%%%%%%%%%%%%%%%%%%%%%%%%%%%%%%%
%%%%%%%%%%%%%%%%%%%%%%%%%%%%%%%%%%%%%%%

In this section, 
we summarize the results obtained in \cite{FMU} and \cite{FMU2}.

%%%%%%%%%%%%%%%%%%%%%%%%%%%%%%%%%%%%%%%
\subsection{Definition of distance}
\label{subsec:definition}
%%%%%%%%%%%%%%%%%%%%%%%%%%%%%%%%%%%%%%%

Let $\mathcal{M}=\{x\}$ be a configuration space 
and $S(x)$ an action. 
We consider a Markov chain for probability distributions $p_n(x)$, 
\begin{align}
 p_n(x)\to p_{n+1}(x)=\int dy\,P(x|y)\,p_n(y),
\end{align}
and design the transition matrix $P(x|y)=\langle x | \hat{P} | y \rangle$ 
such that $P_n(x|y)\equiv \langle x | \hat{P}^n | y \rangle$ 
converges uniquely in the limit $n\to\infty$ 
to the equilibrium distribution 
$p_{\rm eq}(x)\equiv (1/Z)\,e^{-S(x)}$ ($Z=\int dx\, e^{-S(x)}$). 
We further assume that $P(x|y)$ satisfies the detailed balance condition 
\begin{align}
 P(x|y)\,p_{\rm eq}(y) = P(y|x)\,p_{\rm eq}(x),
\end{align}
and that all the eigenvalues of $\hat{P}$ are positive. 
These assumptions are satisfied for Langevin algorithms, 
and if $\hat{P}$ has negative eigenvalues, 
we then instead consider $\hat{P}^2$ as the fundamental transition matrix, 
which satisfies the detailed balance condition of the same form.

Suppose that the system is in equilibrium with $p_{\rm eq}(x)$, 
and consider the set of all $n$-step random paths in $\mathcal{M}$. 
We introduce the {\em connectivity} $f_n(x,y)$ 
between configurations $x$ and $y$ for fixed step number $n$ 
as the ratio of 
``the number of $n$-step paths from $y$ to $x$'' 
to ``the number of all $n$-step paths'' \cite{FMU}.  
This can be expressed as the product of 
the probability to find $y$ in equilibrium 
and the probability to obtain $x$ from $y$ at $n$ steps: 
\begin{align}
 f_n(x,y) = P_n(x|y)\,p_{\rm eq}(y).
\end{align}
The detailed balance condition means that 
$f_n(x,y)$ is a symmetric function of $x$ and $y$. 
We further introduce the {\em normalized connectivity} $F_n(x,y)$ \cite{FMU} 
by 
\begin{align}
 F_n(x,y) \equiv \frac{f_n(x,y)}{\sqrt{f_n(x,x)\,f_n(y,y)}}
 = \sqrt{\frac{P_n(x|y)\,P_n(y|x)}{P_n(x|x)\,P_n(y|y)}}. 
\label{distance}
\end{align}
The {\em distance $d_n(x,y)$ between $x$ and $y$ for fixed $n$ steps} \cite{FMU} 
is then defined by the relation 
$F_n(x,y) = e^{-(1/2)\,d_n^2(x,y)}$,
i.e.\
\begin{align}
 d_n(x,y) \equiv \sqrt{-\ln \frac{P_n(x|y)\,P_n(y|x)}{P_n(x|x)\,P_n(y|y)}}.
\end{align}
One can show that this satisfies all the axioms of distance 
but the triangle inequality \cite{FMU}, 
and that the triangle inequality does hold 
for the coarse-grained configuration space that is introduced below \cite{FMU2}. 
Furthermore, if the Markov chain generates only local moves in $\mathcal{M}$, 
the distance is universal 
in the sense that the difference of distances 
for two different Markov chains with the same action 
can be absorbed into a rescaling of step number $n$. 

For example, the Gaussian action $S(x)=(\beta/2)\,\sum_{i=1}^N x_i^2$ 
for an $N$-dimensional variable $x=(x_i)$ 
(using as a Markov chain the Langevin algorithm 
with fictitious time increment $\varepsilon$) 
gives a flat and translationally invariant geometry \cite{FMU}:
\begin{align}
 d_n(x,y) = \sqrt{\frac{\beta}{2\sinh\beta n\varepsilon}}\,|x-y|.
\end{align}
As another example, 
for the one-dimensional double-well action 
$S(x)=(\beta/2)\,(x^2-1)^2$ with large $\beta$, 
the distance between two minima $x=\pm 1$ can be estimated to be $O(\beta)$ 
\cite{FMU}.

%%%%%%%%%%%%%%%%%%%%%%%%%%%%%%%%%%%%%%%
\subsection{Coarse-grained configuration space}
\label{subsec:course-grain}
%%%%%%%%%%%%%%%%%%%%%%%%%%%%%%%%%%%%%%%

For a Markov chain that generates local moves, 
the distance takes significant values 
only for transitions between configurations around different modes, 
and thus, configurations around the same mode can be effectively treated 
as a single point 
when we discuss about the global geometry of $\mathcal{M}$. 
This leads us to the idea of the {\em coarse-grained configuration space} 
$\bar{\mathcal{M}}$ \cite{FMU}. 
For example, 
for the double-well action above, 
the original configuration space is $\mathcal{M} = \mathbb{R}$, 
and the coarse-grained configuration space consists of 
two minima, $\bar{\mathcal{M}} = \{+1,-1\}$. 
For an $N$-dimensional periodic action 
\begin{align}
 S(x;\beta)=\beta\,\sum_{i=1}^N (1-\cos x_i)~~(\beta\gg 1),
\label{cosine}
\end{align}
the original configuration space is $\mathcal{M}=\mathbb{R}^N$, 
and the coarse-grained configuration space is 
an $N$-dimensional lattice consisting of local minima, 
$\bar{\mathcal{M}}  = \{x=(2\pi k_i) \,|\, k_i\in \mathbb{Z}~(i=1,\ldots,N)\} 
\simeq \mathbb{Z}^N$. 
When an action $S(x)$ has local minima 
that are scattered in the configuration space in a complicated way, 
the gradient flow $\dot{x}^i_\tau = - \partial_i S(x_\tau)$ 
\cite{Luscher:2009eq,Luscher:2010iy}
can be used for a systematic construction of the coarse-grained configuration space.

%%%%%%%%%%%%%%%%%%%%%%%%%%%%%%%%%%%%%%%
\subsection{Geometry for a tempered stochastic process}
\label{subsec:tempering}
%%%%%%%%%%%%%%%%%%%%%%%%%%%%%%%%%%%%%%%

\begin{figure}[htbp]
 \centering
 \includegraphics[width=50mm]{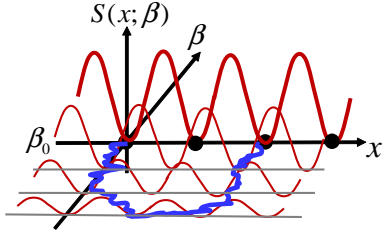} 
 \caption{
 A random path in the extended configuration space $\mathcal{M}\times\mathcal{A}$. 
 $\beta_0$ is the original value of $\beta$, 
 at which the equilibrium distribution is highly multimodal 
 in the $x$-direction. 
 }
 \label{fig:simtemp}
\end{figure}

Suppose that we have a configuration space 
with a highly multimodal equilibrium distribution, 
and let us apply the simulated tempering algorithm \cite{Marinari:1992qd} 
to the system. 
Namely, we treat a parameter $\beta$ existing in the action $S(x;\beta)$ 
(say, the overall coefficient in the action as in \eqref{cosine}) 
as another dynamical variable, 
and extend the configuration space $\mathcal{M}$ 
to $\mathcal{M}\times \mathcal{A} = \{X=(x,\beta)\}$. 
Here $\mathcal{A}=\{\beta\}$ is 
a discrete set of values of $\beta$ 
that includes such values 
for which the distribution $\propto e^{-S(x;\beta)}$ is far less multimodal 
in the $x$-direction. 
We introduce a new Markov chain 
to the extended configuration space $\mathcal{M}\times \mathcal{A}$ 
such that global equilibrium is realized  
with the probability distribution 
$P_{\rm eq}(X)=P_{\rm eq}(x,\beta) \propto e^{-S(x;\beta)}$. 
This algorithm prompts transitions between different modes 
at the original value of $\beta$, 
because they now can communicate with each other 
by passing through configurations at such $\beta$'s that give less multimodality 
(see Fig.~\ref{fig:simtemp}). 

We can also consider the coarse-graining of the extended configuration space. 
In \cite{FMU2}, it is shown for the action \eqref{cosine} 
that the coarse-grained, extended configuration space 
$\bar{\mathcal{M}}\times\mathcal{A}$ 
has a geometry of Euclidean AdS$_{N+1}$ of the following form for large $\beta$: 
\begin{align}
 ds^2 &\equiv d_n^2(X,X+dX) = {\rm const.}\,\beta^q\,\sum_{i=1}^N dx_i^2 
 + {\rm const.}\,\frac{d\beta^2}{\beta^2},
\label{AdS}
\end{align}
which can be transformed to a standard form, 
$ds^2 \propto \rho^2 \sum_{i=1}^N dx_i^2 + d\rho^2/\rho^2
\propto (\sum_{i=1}^N dx_i^2 + dz^2)/z^2$, 
by setting $\beta = {\rm const.}\, \rho^{2/q} = {\rm const.}\, z^{-2/q}$. 

We comment that the discretization of $\mathcal{A}$ must be done 
so that configurations can move smoothly in the $\beta$-direction. 
A simple, geometrical analysis based on the distance shows that 
an exponential stepping is optimal for large $\beta$ \cite{FMU2,Fukuma:2020got}, 
and this result was used in the tempered Lefschetz thimble method 
\cite{Fukuma:2017fjq,Fukuma:2019wbv,Fukuma:2019uot,Fukuma:2020lny}, 
that is a method towards solving the numerical sign problem.

%%%%%%%%%%%%%%%%%%%%%%%%%%%%%%%%%%%%%%%
%%%%%%%%%%%%%%%%%%%%%%%%%%%%%%%%%%%%%%%
\section{Geometry for a stochastic process of matrices}
\label{sec:matrix}
%%%%%%%%%%%%%%%%%%%%%%%%%%%%%%%%%%%%%%%
%%%%%%%%%%%%%%%%%%%%%%%%%%%%%%%%%%%%%%%

%%%%%%%%%%%%%%%%%%%%%%%%%%%%%%%%%%%%%%%
\subsection{$U(N)$ matrix model}
\label{subsec:matrix_model}
%%%%%%%%%%%%%%%%%%%%%%%%%%%%%%%%%%%%%%%

The appearance of asymptotic AdS$_{N+1}$ metric 
may not be so much interesting 
as a model of gravity, 
because in quantum field theories, 
$x=(x_i)$ corresponds to a configuration of field variables 
(such as a configuration of link variables on the lattice 
for the lattice gauge theory), 
and thus the degrees of freedom, $N$, is infinite 
in the thermodynamic limit. 
It is also hard to regard the configuration space  
as directly related to a spacetime.

However, things become different 
if we note that \eqref{cosine} is a one-body part of the $U(N)$ matrix model, 
\begin{align}
 S(U;\beta) = \beta N\,{\rm Re}\,{\rm tr}\,(1-U)
 =\frac{\beta N}{2}\,{\rm tr}\,(2-U-U^\dag),
\end{align}
with $U$ set to a diagonal form $U={\rm diag}\,[i x_j]$ $(j=1,\ldots,N)$, 
and if we recall that the eigenvalue can be regarded as a spacetime coordinate 
(see, e.g., \cite{Das:1990kaa}). 
To state this more precisely, 
we first rewrite the partition function  
as the integration over eigenvalues:
\begin{align}
 Z \equiv \int dU\,e^{-S(U;\beta)}
 \propto  \int_{-\pi}^\pi \Bigl(\prod_{i=1}^N \frac{dx_i}{2\pi}\Bigr)\,
 \Bigl(\prod_{i<j} \sin^2 \frac{x_i-x_j}{2}\Bigr)\,
 e^{-\beta N\,\sum_{i=1}^N (1-\cos x_i)}.
\end{align}
The Faddeev-Popov determinant $\prod_{i<j} \sin^2 (x_i-x_j)/2$ 
gives a repulsive two-body potential between eigenvalues. 
Then, by introducing the collective coordinate $u$ through 
\begin{align}
 \hat\rho (u) = \frac{1}{N}\,\sum_{i=1}^N \delta_P(x_i-u)
 ~~(\mbox{$\delta_P(x)$: periodic delta function with period $2\pi$}),
\end{align}
we now can regard $u$ as a coordinate 
of a covering space with a period $2\pi m$, 
$\mathcal{M}=\{u|-\pi m\leq u<\pi m\}$.

It is known that the matrix model has a third-order phase transition 
in the large $N$ limit at the Gross-Witten-Wadia point $\beta=1$ 
\cite{Gross:1980he,Wadia:1980cp}, 
where the functional form of the eigenvalue distribution
$\rho(u)\equiv \lim_{N\to\infty} \langle \hat\rho(u) \rangle$ 
changes as follows 
($x_c\equiv 2\arcsin(1/\sqrt{\beta})$):
\begin{itemize}
\item
$\beta\geq 1$:\vspace{-7mm}
\begin{align}
 \rho(u)=\left\{ 
  \begin{array}{cl}
  \dfrac{\beta}{\pi}\,\cos\dfrac{u}{2}\cdot 
  \sqrt{\sin^2\dfrac{x_c}{2}-\sin^2\dfrac{u}{2}}
  & ~~~(u\in I_1) \\
  0
  & ~~~(u\in I_2) 
  \end{array}
 \right.
\label{density_low_temp}
\end{align}
\item
$\beta\leq 1$:\vspace{-7mm}
\begin{align}
 \rho(u) = \frac{1}{2\pi}\,(1+\beta\cos u)~~~(\forall u),
\label{density_high_temp}
\end{align}
\end{itemize}
where $I_1\equiv \{u \,| -x_c \leq [u] \leq x_c\}$ 
and 
$I_2\equiv \{u \,| -\pi \leq [u] \leq -x_c\,\,\mbox{or} +x_c\leq [u] < \pi\}$ 
with $[u]$ the projected value of $u$ to the fundamental region $[-\pi,\pi)$.

For a finite but large $N$, 
one-eigenvalue transitions yield an instanton effect of $e^{-O(N)}$ 
in the low-temperature phase ($\beta\geq 1$) \cite{David:1990sk} 
(see, e.g., \cite{Marino:2008ya,Ahmed:2017lhl} 
for nonperturbative effects of the Gross-Witten-Wadia model). 
One can easily show that the one-eigenvalue feels 
the effective potential $S_{\rm eff}(u;\beta) = N\,v(u;\beta)+O(\ln N)$ with
\begin{align}
 v(u) \equiv \left\{
 \begin{array}{cl}
  0
  & (u\in I_1) \\
  2\,\Bigl|\dfrac{\sin u/2}{\sin x_c/2}\Bigr|\,
  \sqrt{\dfrac{\sin^2 u/2}{\sin^2 x_c/2}-1}
  -2\,\ln\biggl( \Bigl|\dfrac{\sin u/2}{\sin x_c/2}\Bigr|
  + \sqrt{\dfrac{\sin^2 u/2}{\sin^2 x_c/2}-1}\biggr)
  & (u\in I_2).
 \end{array}
 \right.
\end{align}
The vanishing of potential in the region $|u|\leq x_c$ 
(we have set a possible additive constant to zero)
can be understood 
as a balance between the one-body potential and the repulsive potential 
from other eigenvalues that are condensed 
with the distribution \eqref{density_low_temp}. 
Note that there is no such instanton effect for the high temperature phase, 
so that $v(u)$ vanishes for $\beta\leq 1$.

%%%%%%%%%%%%%%%%%%%%%%%%%%%%%%%%%%%%%%%
\subsection{Geometry for a stochastic process of one-eigenvalue and a string theoretical interpretation}
\label{subsec:geometry_one-eigenvalue}
%%%%%%%%%%%%%%%%%%%%%%%%%%%%%%%%%%%%%%%

Let us consider the stochastic process of one-eigenvalue $u$ 
with respect to the effective action 
$S(u;\beta)=N\,v(u;\beta) + O(\ln N)$ for $\beta\geq 1$. 
The distance measures the difficulty of one-eigenvalue transitions 
in the background where other eigenvalues are condensed 
to large-$N$ solutions \eqref{density_low_temp}. 
We thus can interpret the distance 
as ``measuring the background geometry of condensed eigenvalues 
by using the one-eigenvalue as a probe.'' 
On the other hand, in noncritical string theory, 
one-eigenvalue corresponds to a D-instanton \cite{Fukuma:1999tj} 
(see also \cite{Fukuma:1996hj,Fukuma:1996bq,Fukuma:2005nm,Fukuma:2006qq,
Fukuma:2006ny,Fukuma:2007qz}). 
Thus, the above can be rephrased 
as ``measuring the background geometry 
by using a D-instanton as a probe.''

%%%%%%%%%%%%%%%%%%%%%%%%%%%%%%%%%%%%%%%
\subsection{Emergence of a quantum blackhole}
\label{subsec:blackhole}
%%%%%%%%%%%%%%%%%%%%%%%%%%%%%%%%%%%%%%%

We now implement the simulated tempering 
by treating the inverse 't Hooft coupling $\beta$ as a random variable. 
The coarse-grained, extended configuration space 
$\bar{\mathcal{M}}\times\mathcal{A}=\{X=(u,\beta)\}$ 
then becomes a two-dimensional lattice. 
Recall that we have set a periodic boundary condition for the $u$-direction 
with period $2\pi m$. 

\begin{figure}[htbp]
 \centering
 \includegraphics[width=50mm]{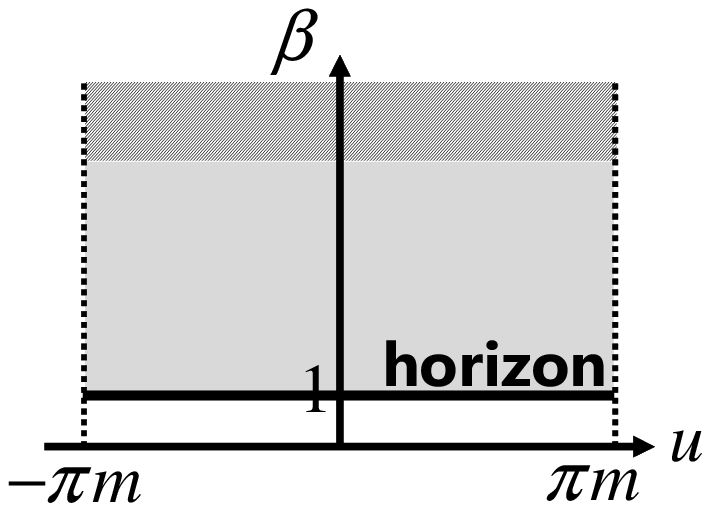} \hspace{3mm}
 \includegraphics[width=35mm]{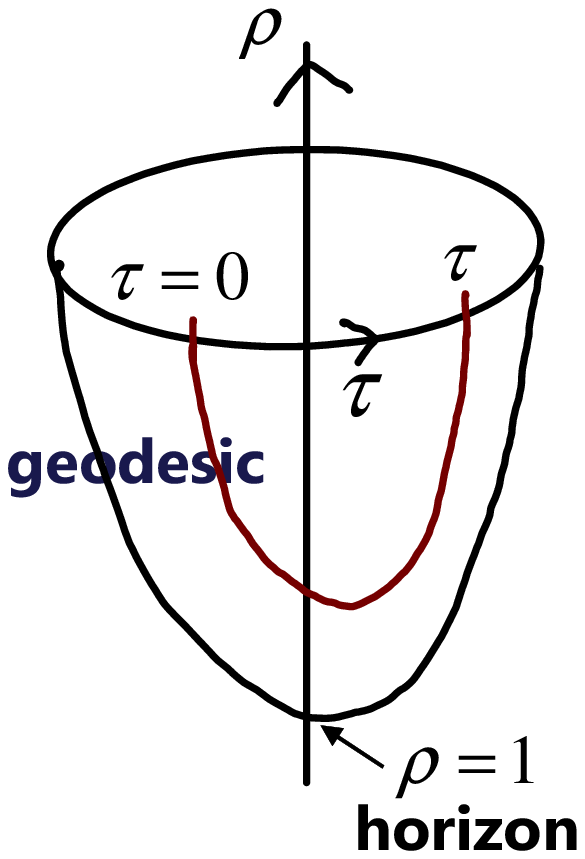}
 \caption{
 (Left) Geometry of $\bar{\mathcal{M}}\times\mathcal{A}$ 
 in the large-$N$ limit. 
 $u$ has a period $2\pi m$,  
 and the $S^1$-cycle vanishes at the horizon $\beta=1$. 
 (Right) The corresponding geometry in continuum. 
 $\tau$ has a period $2\pi$, and the horizon is at $\rho=1$. 
 }
 \label{fig:AdS2}
\end{figure}

The geometry of $\bar{\mathcal{M}}\times\mathcal{A}$ 
in the low temperature phase ($\beta\geq 1$) 
should be as follows. 
First we recall that the effective potential disappears at $\beta=1$ 
in the large-$N$ limit, 
which means that the distance between $(0,\beta)$ and $(u,\beta)$ 
should become vanishingly small at $\beta=1$ for arbitrary $u$. 
Since we set a periodic boundary condition in the $u$-direction, 
we thus find that the $S^1$-cycle in the $u$-direction vanishes at $\beta=1$ 
in the large-$N$ limit (see the left panel of Fig.~\ref{fig:AdS2}). 
On the other hand, 
the effective action $S_{\rm eff}(u;\beta)=N\,v(u;\beta) + O(\ln N)$ 
has many local minima for $\beta\gg 1$, 
which means that the geometry of $\bar{\mathcal{M}}\times\mathcal{A}$ 
must be asymptotically AdS$_2$ for large $\beta$. 
Combining these two observations, 
we expect that 
the full geometry of $\bar{\mathcal{M}}\times\mathcal{A}$ for $\beta\geq 1$ 
is given by a Euclidean AdS$_2$ space 
with a horizon at the Gross-Witten-Wadia point, $\beta=1$. 
We choose a candidate metric in continuum of the form
\begin{align}
 ds^2 = \ell^2\Bigl[ (\rho^2-1)\,d\tau^2 + \frac{d\rho^2}{\rho^2-1}\Bigr],
\label{AdS2BH}
\end{align}
which is defined only for the outside of the horizon, $\rho\geq 1$ 
(see the right panel of Fig.~\ref{fig:AdS2}). 
One can easily check that the metric \eqref{AdS2BH} satisfies 
the equations $R_{\mu\nu}=\Lambda\, g_{\mu\nu}$ with $\Lambda=-1$ 
and has no conical singularity at the horizon $\rho=1$ 
when $\tau$ has a period $2\pi$. 
Furthermore, the geodesic distance $I(\tau,\rho)$ 
between two points $(0,\rho)$ and $(\tau,\rho)$ 
can be analytically calculated to be
\begin{align}
 I(\tau,\rho) = 2\ell \,{\rm arccosh}
 \sqrt{\cos^2\frac{\tau}{2}+\rho^2\sin^2\frac{\tau}{2}}.
\label{geodesic}
\end{align}
In the following discussions, 
we will set an ansatz that the two coordinate systems $(u,\beta)$ and $(\tau,\rho)$ 
are related to each other with a simple scaling: 
\begin{align}
 u=m\tau,\quad \rho=\rho_c + \alpha (\beta-\beta_c)^{q/2}.
\label{relation}
\end{align}

%%%%%%%%%%%%%%%%%%%%%%%%%%%%%%%%%%%%%%%
\subsection{Numerical confirmation}
\label{subsec:numerical}
%%%%%%%%%%%%%%%%%%%%%%%%%%%%%%%%%%%%%%%

We now numerically confirm that the two geometries in Fig.~\ref{fig:AdS2} 
are the same under the relation \eqref{relation}, 
by comparing ``the numerically obtained distances 
$d_n(X,Y)$ between $X=(0,\beta)$ and $Y=(u,\beta)$'' 
with ``the analytic values of distance, $I(\tau,\rho)$, 
between $(0,\rho)$ and $(\tau,\rho)$'' 
for various $(u,\beta)$ 
with $u=u_j=2\pi j$ $(j=1,\ldots,25)$ 
and $\beta=\beta_a=5\times 5^{-a/5}$ $(a=0,1,\ldots,5)$.

\begin{figure}[htbp]
 \includegraphics[width=45mm]{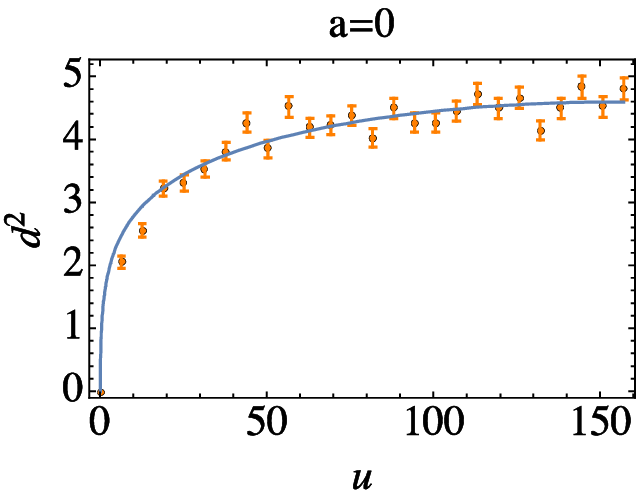} \hspace{3mm}
 \includegraphics[width=45mm]{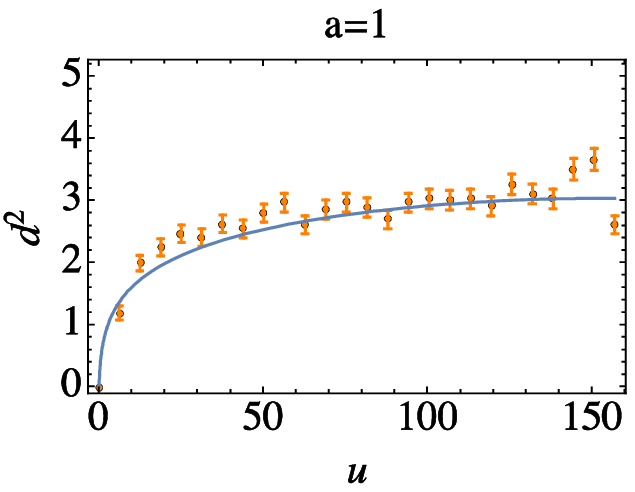} \hspace{3mm}
 \includegraphics[width=45mm]{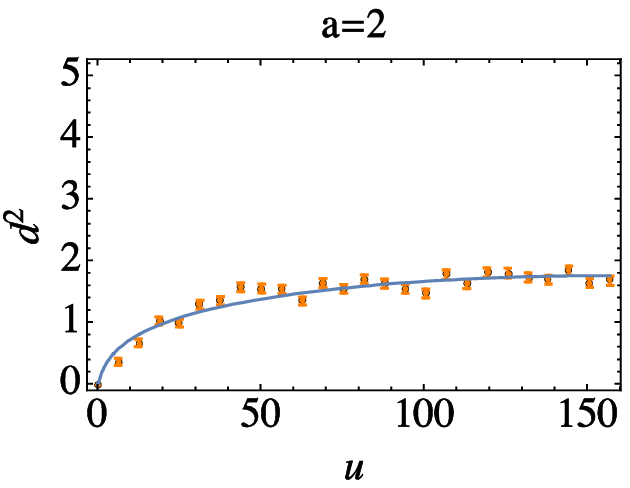} \\
 \vspace{3mm}
 \includegraphics[width=45mm]{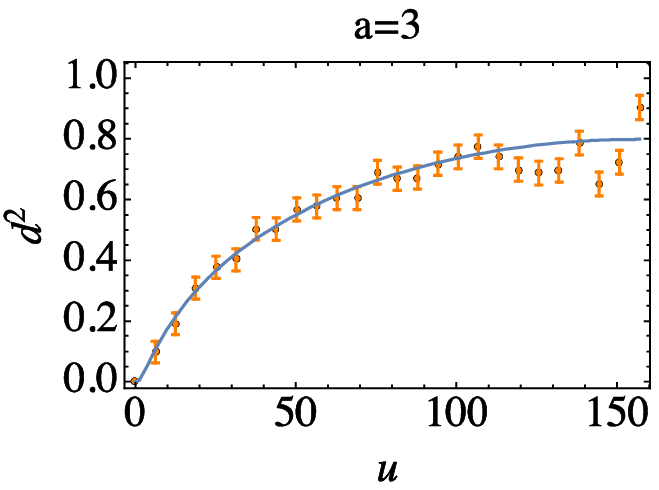} \hspace{3mm}
 \includegraphics[width=45mm]{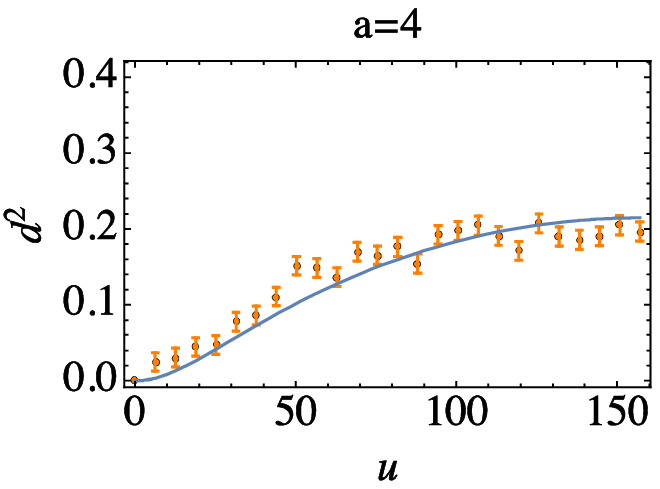} \hspace{3mm}
 \includegraphics[width=45mm]{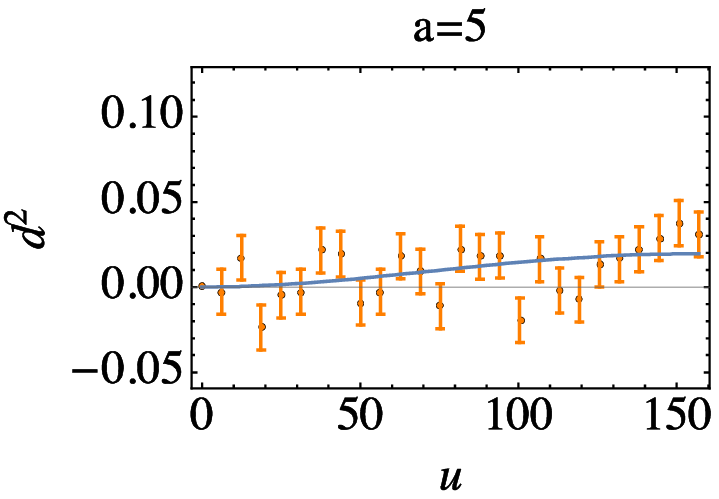} \\
 \caption{
 Numerically obtained squared distances $d_n^2(X,Y)$. 
 The solid lines represent the squared geodesic distances, 
 $I^2(\tau,\rho)$, 
%  for the optimizing parameters \eqref{optimizing}.
 for the optimizing parameters (3.11).
}
 \label{fig:geodesic}
\end{figure}

With $N=20$, $n=200$ and $m=50$, 
we determine the parameters $(\ell,\beta_c,\alpha,\rho_c,q)$ 
by minimizing 
\begin{align}
 \chi^2 %(\ell,\beta_c,\alpha,\rho_c,q)
 \equiv
 \sum_j \sum_a \Bigl[\frac{d_n^2((0,\beta_a),
 (u_j,\beta_a))-I^2(\tau(u_j),\rho(\beta_a))}
 {\delta d_n^2((0,\beta_a),(u_j,\beta_a))}\Bigr]^2,
\end{align}
where $[\delta d_n^2(X,Y)]^2$ is a sample variance in the estimation of $d_n^2(X,Y)$. 
The obtained results are shown in Fig.~\ref{fig:geodesic} 
for the optimizing values, 
\begin{align}
 \ell=0.10,~\beta_c=0.33,~\alpha=3.0,~\rho_c=0.93,~q=11
\label{optimizing}
\end{align}
with $\sqrt{\chi^2/{\rm DOF}}=\sqrt{\chi^2/(25\times 6-5)}=1.6$, 
from which we confirm that 
the geometry of $\bar{\mathcal{M}}\times\mathcal{A}$ 
is well given by the metric \eqref{AdS2BH} 
through the relation \eqref{relation}. 

We comment that, for the optimizing parameters \eqref{optimizing}, 
the corresponding value of $\beta$ to the horizon $\rho=1$ 
is given by $\beta=0.83$. 
We expect that the discrepancy from the critical value $\beta=1$ 
can be understood as a finite $N$ correction 
(or it may imply that the relation \eqref{relation} needs to be modified) 
\cite{FM2}. 

%%%%%%%%%%%%%%%%%%%%%%%%%%%%%%%%%%%%%%%
%%%%%%%%%%%%%%%%%%%%%%%%%%%%%%%%%%%%%%%
\section{Conclusion and outlook}
\label{sec:conclusion}
%%%%%%%%%%%%%%%%%%%%%%%%%%%%%%%%%%%%%%%
%%%%%%%%%%%%%%%%%%%%%%%%%%%%%%%%%%%%%%%

In this article, 
we applied the distance between configurations, \eqref{distance},  
to a Markov process of matrices. 
By identifying the eigenvalues with coordinates of a spacetime, 
this realizes a mechanism for 
the {\em emergence of spacetime geometry from randomness}. 
As an example, we showed that there emerges a Euclidean AdS geometry 
with a horizon from a tempered stochastic process of $U(N)$ matrix. 

There are left many things we need to study in a more elaborate way. 
First of all, the fitting of the geometry with metric \eqref{AdS2BH} 
is based on the assumption \eqref{relation}. 
It should be nice if we can analytically calculate the distance $d_n(X,Y)$ 
by using the tempered Langevin algorithm 
for the one-eigenvalue effective action $S_{\rm eff}(u;\beta)$. 
It should also be interesting to understand the results obtained here 
in the context of AdS/CFT correspondence, 
especially with a relation to the SYK model 
\cite{Sachdev:1992fk,Kitaev2015} and the wormholes.

It must be important to generalize the present results to higher dimensions. 
We expect that an AdS$_{d+1}$ blackhole will emerge 
from a tempered stochastic process of 
the $d$-dimensional twisted Eguchi-Kawai model 
\cite{GonzalezArroyo:1982hz}, 
where the action is given by $S(U_\mu;\beta)=\beta \,{\rm Re}\,{\rm tr}\, 
(1-z_{\mu\nu}U_\mu U_\nu U_\mu^\dag U_\nu^\dag)$ 
for $U_\mu\in U(N)$ $(\mu=1,\ldots,d)$. 
The spacetime coordinates $x=(x_\mu)\in\mathbb{R}^d$ can be identified with 
the fluctuating modes corresponding to $U(1)^d$ transformations 
$U_\mu \to e^{i\,x_\mu}\,U_\mu$. 
For a symmetric twist, 
this $U(1)^d$ symmetry is actually broken at large $\beta$ 
if $N$ is taken to be sufficiently large \cite{Teper:2006sp}. 
In fact, as was also first found in \cite{Teper:2006sp}, 
the $U(1)^d$ symmetry is restored step by step 
(the number of restored directions increases one by one 
as $\beta$ decreases). 
Since in our construction 
the spacetime can be extended only in the {\em broken} directions 
and the $S^1$-cycles vanish for the restored directions, 
we expect that an AdS$_{d+1}$ blackhole with a single horizon 
is obtained for the region $\beta\geq \beta_H$, 
where $\beta_H$ corresponds to the coupling 
at which a $U(1)$ symmetry is restored for the first time 
when reducing $\beta$ from a sufficiently large value. 

As another direction of future project, 
it should be important to investigate 
whether we can find a stochastic process that gives de Sitter space. 
If such a process exists, 
then this should give a new mechanism to 
{\em create an inflationary universe from randomness}. 

A study along these lines is now in progress and will be reported elsewhere. 

%%%%%%%%%%%%%%%%%%%%%%%%%%%%%%%%%%%%%%%
%%%%%%%%%%%%%%%%%%%%%%%%%%%%%%%%%%%%%%%
\section*{Acknowledgments}
It is a pleasure to thank the organizers of Corfu 2019, 
especially George Zoupanos and Konstantinos Anagnostopoulos, 
for their very warm hospitality offered to one of us (M.F.). 
We also thank Gerald Dunne, Kenichi Ishikawa, Hikaru Kawai, 
Holger Nielsen, Shinsuke Nishigaki, 
Jun Nishimura, Asato Tsuchiya 
and especially Naoya Umeda 
for enlightening discussions. 
This work was partially supported by JSPS KAKENHI 
(Grant Numbers 16K05321, 18J22698)   
and by SPIRITS 2019 of Kyoto University (PI: M.F.).
%%%%%%%%%%%%%%%%%%%%%%%%%%%%%%%%%%%%%%%
%%%%%%%%%%%%%%%%%%%%%%%%%%%%%%%%%%%%%%%

\end{document}